# Measurement-based estimation of global pupil functions in 3D localization microscopy

Petar N. Petrov,[1] Yoav Shechtman,[1,2] and W. E. Moerner[1,*]

[1]*Department of Chemistry, Stanford University, 333 Campus Drive, Stanford, CA 94305, USA*
[2]*Present address:\ Department of Biomedical Engineering, Technion−Israel Institute of Technology, Haifa 32000, Israel*
*[\*wmoerner@stanford.edu](mailto:wmoerner@stanford.edu)*

**Abstract:** We report the use of a phase retrieval procedure based on maximum likelihood estimation (MLE) to produce an improved, experimentally calibrated model of a point spread function (PSF) for use in three-dimensional (3D) localization microscopy experiments. The method estimates a global pupil phase function (which includes both the PSF and system aberrations) over the full axial range from a simple calibration scan. The pupil function is used to refine the PSF model and hence enable superior localizations from experimental data. To demonstrate the utility of the procedure, we apply it to experimental data acquired with a microscope employing a tetrapod PSF with a 6 µm axial range. The phase-retrieved model demonstrates significant improvements in both accuracy and precision of 3D localizations relative to the model based on scalar diffraction theory. The localization precision of the phase-retrieved model is shown to be near the limits imposed by estimation theory, and the reproducibility of the procedure is characterized and discussed. Code which performs the phase retrieval algorithm is provided.

### 1. Introduction

High-precision localization of nanoscale fluorescent emitters such as quantum dots, nanoparticles, fluorescent beads, and single molecules has enabled optical imaging techniques to track single particles [1-3] and visualize sub-diffraction structures in great detail and with molecular specificity via superresolution imaging [4-7]. Historically, such localization microscopy has been widely used to acquire two-dimensional position data, in which the lateral $(x, y)$ coordinates of an emitter are determined from its point spread function (PSF), namely, the image a point source creates on the camera, via a centroid calculation or a two-dimensional fit to a model function (e.g. Gaussian) [8].

In principle, this procedure can be extended into three dimensions by taking advantage of the dependence of the shape of the PSF on the axial $(z)$ position of the emitter. However, the PSF of a standard microscope is not suited to three-dimensional (3D) localization microscopy, as its pattern is approximately symmetric about the focal plane and spreads out rapidly on the detector as the image is defocused. In order to circumvent these shortcomings, various engineered PSFs have been developed using Fourier (pupil) plane processing. By phase-modulating the collected light in the Fourier plane of the microscope, one can change the shape of the PSF to provide more information about axial position. Examples of this include the astigmatic [9], double-helix [10,11], corkscrew [12], and self-bending [13] PSFs. Recently, our lab demonstrated another family of PSFs called tetrapods, which not only extend the axial range over which localizations can be achieved to as much as 20 µm, but are able to deliver optimal 3D localization precision over a chosen axial range by encoding the maximum amount of position information in their shape [14,15].

While the tetrapod PSFs are optimal for the problem of 3D localization microscopy in terms of information content given fixed signal photons and background, the complexity of their shapes poses an analytical challenge. Whereas many engineered PSFs can be parameterized in terms of simple functions like 2D Gaussians [9,10,12,13], the shapes formed by the tetrapod PSFs are too intricate and rapidly-axially-varying for this approach, and as such require a more detailed model in order to be utilized effectively. Models based on calculations of the theoretical electromagnetic field in the pupil plane of the microscope, or the pupil function, have been demonstrated successfully for the tetrapod PSF [15,16].

Unfortunately, the performance of the theoretical model can be hampered by the presence of optical aberrations, which are a pervasive feature of any practical imaging system. These aberrations introduce undesirable phase modulation into the microscope response, resulting in distortions of the PSF shape which are not captured by the theoretical model. As we show below, even subtle distortions can significantly degrade the 3D localization performance when employing a theoretical model [17].

Previously, the problem of optical aberrations in 3D localization microscopy has usually been addressed by correcting aberrations using adaptive optics [18] or improving the PSF model (over ~1 μm in $z$) using a phase retrieval-based pupil function characterization [19,20] based on the modified Gerchberg-Saxton algorithm presented in [21]. However, for PSFs with large axial ranges, it has been found [14,17] that the algorithm must be used locally at multiple regions throughout the axial range of the PSF, requiring multiple pupil functions to describe the system. There is a need for an easy-to-use procedure that avoids the complexity of multiple local pupil functions.

Here we demonstrate an alternative, global method of phase retrieval which extracts the pupil function of the microscope directly from a set of experimentally-acquired PSF images and produces a measurement-based imaging model which restores the sub-diffraction accuracy and precision of the localization process using a single pupil function. We illustrate the algorithm for the 6 µm tetrapod PSF and show that one pupil function works over its entire 6 µm axial range. Compared to multiple local pupil functions, this method is more elegant, more theoretically rigorous, and extendable to the case of index mismatch if required.

The phase retrieval procedure begins with an uncorrected theoretical model of the pupil function in an aberration-free optical system with an engineered PSF phase mask added in the pupil plane. Analogous to previous work on the standard open-aperture PSF [22], a phase aberration term, decomposed into the Zernike polynomial basis, is optimized via maximum likelihood estimation (MLE) using the experimentally-measured single-emitter PSF as a benchmark. The resulting estimate of the pupil function of the microscope is used to produce a new, corrected model of the engineered PSF which now accounts for experimentally-observed aberrations in its shape via the phase-retrieved aberration term in the pupil function. For the case of the 6 μm tetrapod PSF, the phase-retrieved PSF model is shown to improve both the precision and accuracy of single-emitter localizations relative to the purely theoretical, uncorrected model. The precision of both the lateral and axial localizations approaches the theoretical limit imposed by the Cramér-Rao lower bound (CRLB) [23-25] across the entire 6 μm axial range.

## 2. Imaging model

### 2.1 Theoretical pupil function

In the scalar diffraction approximation, the PSF in the image plane $I(x',y'|x_o,y_o,z_o)$ due to a monochromatic point source at position $(x_o,y_o,z_o)$ in the object plane satisfies [26]

$$I(x',y'|x_o,y_o,z_o) \propto \left|\mathcal{F}\{E_{FP}(\rho,\varphi|x_o,y_o,z_o)\}\right|^2 \tag{1}$$

where $E_{FP}(\rho,\varphi|x_o,y_o,z_o)$ is the electric field at the Fourier (pupil) plane of the imaging system, given in polar coordinates, and $\mathcal{F}\{\cdot\}$ denotes the two-dimensional spatial Fourier transform.

Due to the presence of optical aberrations, the exact form of the electric field $E_{FP}$ is not known in an experiment. However, in an aberration-free imaging system, we take the theoretical form of the pupil function for an emitter along the optical axis and in the focal plane of the objective ($z_o = 0$) in the absence of refractive index mismatch to be

$$E_0(\rho,\varphi) \equiv E_{FP}(\rho,\varphi\,|\,0,0,0) = \frac{\mathrm{circ}(\rho)}{\left[1-\left(\frac{NA}{n}\rho\right)^2\right]^{1/4}} \exp\left[i\mathfrak{M}(\rho,\varphi)\right] \qquad (2)$$

where $NA$ is the numerical aperture of the imaging system, $n$ is the refractive index of the immersion medium, $\mathrm{circ}(\rho) = \begin{cases} 1, & \rho \leq 1 \\ 0, & \rho > 1 \end{cases}$, and $\mathfrak{M}(\rho,\varphi)$ corresponds to the phase pattern imparted by the phase mask in the Fourier plane of the system. It is this phase pattern (produced by a spatial light modulator or a phase plate) that defines a desired engineered PSF. The radial coordinate, $\rho$, in the Fourier plane is normalized such that its value is unity at the radius of the limiting aperture.

In our microscope, a 4$f$ system is added to the emission path (see experimental details in Section 5) in order to introduce the phase mask in the Fourier plane. As a result, based on the Abbe sine condition, the physical dimension of the limiting radius is given by $f_{4f} NA / \sqrt{M^2 - NA^2}$, where $f_{4f}$ is the focal length of each 4$f$ lens and $M$ is the magnification of the microscope. The denominator in Eq. (2) is an apodization factor which describes the amount of area from the spherical wavefront that is projected onto a unit area in the Fourier plane [27]. This factor leads to a decrease in the amplitude of the pupil function with increasing spatial frequency up to the limiting spatial frequency within the passband of the imaging system.

For objects with a lateral displacement $(x_o, y_o)$ away from the optical axis, a linear phase is applied in the Fourier plane to translate the model PSF in the image plane. This phase shift takes the form

$$\Phi_{lat}(\xi,\eta\,|\,x_o,y_o) = \frac{2\pi}{\lambda f_{4f}}(x_o\xi + y_o\eta) \qquad (3)$$

where $\xi$ and $\eta$ are the horizontal and vertical coordinates in the Fourier plane and $\lambda$ is the emission wavelength of the point source [26]. These coordinates can be related to the normalized polar coordinate system presented above by the transformations $\xi = \left(f_{4f} NA / \sqrt{M^2 - NA^2}\right)\rho\cos\varphi$ and $\eta = \left(f_{4f} NA / \sqrt{M^2 - NA^2}\right)\rho\sin\varphi$, so that phase shifts arising from lateral displacements can be calculated from

$$\Phi_{lat}(\rho,\varphi\,|\,x_o,y_o) = \frac{2\pi NA}{\lambda\sqrt{M^2 - NA^2}}\rho(x_o\cos\varphi + y_o\sin\varphi). \qquad (4)$$

Axial shifts can be considered in two ways: displacement of the emitter away from the focal plane and displacement of the focal plane from the emitter. This distinction is important in the case of a mismatch between the refractive indices of the immersion medium and the sample medium, since the distance of the emitter from the refractive index boundary must be accounted for in calculating the overall electric field [28], as well as possible contributions of supercritical fluorescence emission to the pupil function for emitters very close to the boundary [19,29]. However, here we will exclusively treat the case of index-matched media. In the absence of a refractive index boundary, the phase lag incurred along a ray by propagating the field along the optical axis is parameterized by a single term, $z_o$, equal to the axial displacement between the emitter and the focal plane [30]. This phase shift term due to axial displacement takes the form

$$\Phi_{ax}(\rho,\varphi\,|\,z_o) = \frac{2\pi n}{\lambda} z_o \sqrt{1-\left(\frac{NA}{n}\rho\right)^2} \qquad (5)$$

where positive $z_o$ describes an emitter position away from the objective, and $n$ is assumed to be the refractive index of the cover-glass and index-matched oil ($n \sim 1.518$). The final expression for the theoretical electric field in the Fourier plane is given by

$$E_{FP}(\rho,\varphi \mid x_o, y_o, z_o) = E_0(\rho,\varphi) \exp\{i[\Phi_{lat}(\rho,\varphi \mid x_o, y_o) + \Phi_{ax}(\rho,\varphi \mid z_o)]\}. \qquad (6)$$

With the completed theoretical Fourier plane electric field in hand, the corresponding engineered PSF, $I(x',y' \mid x_o, y_o, z_o)$, is calculated via Eq. (1). The PSF can then be pixelated via detector pixel-size integration of the image plane intensity; however, in practice image interpolation provides similar results and is computationally much faster. Finally, the image is scaled so that its integral matches the total number of signal photons, $N$, and a constant background of $b$ photons per pixel is added to the PSF model to produce the image expected on the detector. Ultimately, for a given electric field such as the one derived above, the model of the engineered PSF is parameterized in terms of the five variables $\boldsymbol{\theta} = [x_o, y_o, z_o, N, b]$, and returns the expected value of the photon counts in each pixel when those five parameters are set.

## 2.2 Incorporation of experimental conditions

Our goal in this paper is to describe a phase-retrieval algorithm that estimates pupil phase aberration in the most useful and direct way possible, and a flowchart of the overall procedure is shown in Fig. 1(a). The algorithm requires an appropriate (aberration-free) theoretical model as described above. Our approach in creating the theoretical imaging model is to begin with a good approximation of the pupil function. Note that since our phase retrieval algorithm finds a local solution (in Zernike space) to the optimization problem, starting from a good initial point is important. In order to best approximate the pupil function in the absence of phase aberrations, we incorporated two key features of our experiment into the theoretical model.

The first such feature is the experimental phase mask pattern. For clarity, we will now specialize to the 6 µm tetrapod mask, one of several phase patterns in the tetrapod family, which generates a PSF with a 6 µm axial range. Although the optimal 6 µm tetrapod phase pattern [15] (not shown) is a continuous function, the lithographically fabricated quartz phase mask that was placed in our experimental setup is a discretized version of the continuous pattern. This mask was designed to operate only at a single wavelength (660 nm or 550 nm), which is in general not the emission wavelength of the fluorescent beads used in the demonstration experiments of this paper. In order to generate an appropriate $\mathfrak{M}(\rho,\varphi)$ we use a blueprint of the quartz phase mask pattern, in units of physical quartz thickness, and convert it to a phase term by computing the optical path length through each pixel of the mask at the mean emission detection wavelengths of the beads of interest here (610 nm or 522 nm, see Section 5).

The second essential feature of the model is a discrepancy between the diameter of the phase mask (2.65 mm) and the diameter of the Fourier plane electric field (2.52 mm). This difference results in a slight clipping of the phase pattern by the electric field, affecting the phase delay experienced by the different k-vectors in the field and altering the observed PSF.

Incorporating both of the above features into the theoretical pupil function calculation generates a theoretical PSF model which is a good initial point for the phase retrieval algorithm.

## 2.3 Phase aberration component

While the theoretical PSF based on the model presented above is in good qualitative agreement with the experimental tetrapod PSF, small discrepancies such as those shown in Fig. 2 are often observed. Unfortunately, even small discrepancies over parts of the axial range can limit the accuracy and precision of localizations, which is why we seek a solution over the entire range. We are now ready to improve the theoretical PSF by including additional information from an experimental calibration scan by adding a phase aberration term to the theoretical pupil function. We choose to represent the additional phase aberration in terms of Zernike polynomials, which form a suitable, compact basis for common optical aberrations [30,31]. This term takes the form

$$\Psi(\rho,\varphi\,|\,\boldsymbol{c}) = \sum_{j=4}^{J} c_j Z_j(\rho,\varphi) \quad (7)$$

where $Z_j(\rho,\varphi)$ is the Zernike polynomial with Noll index [32] $j$ and $\boldsymbol{c} = [c_4, c_5, \ldots, c_J]$ is a real vector. This single phase aberration term, $\Psi$, is then incorporated into the field to produce the global phase-aberrated pupil function

$$E_{FP}^{\Psi}(\rho,\varphi\,|\,0,0,0;\boldsymbol{c}) = E_{FP}(\rho,\varphi\,|\,0,0,0)\exp[i\Psi(\rho,\varphi\,|\,\boldsymbol{c})]. \quad (8)$$

Our goal is to determine the values of the polynomial coefficients, $c_j$, such that addition of the phase aberration term to the theoretical electric field in the Fourier plane produces an improved PSF model which more closely matches the experimentally observed PSF. In turn, the improvements in the model lead to higher accuracy and precision of the emitter localization procedure. We choose the Zernike basis for expansion of this additional term because of the similarity of the low order Zernike polynomials to commonly observed optical aberrations [30,31], but this choice is not fundamental to the procedure. Furthermore, the summation excludes the first three Zernike polynomials (piston, tip, and tilt) because they do not change the shape of the PSF.

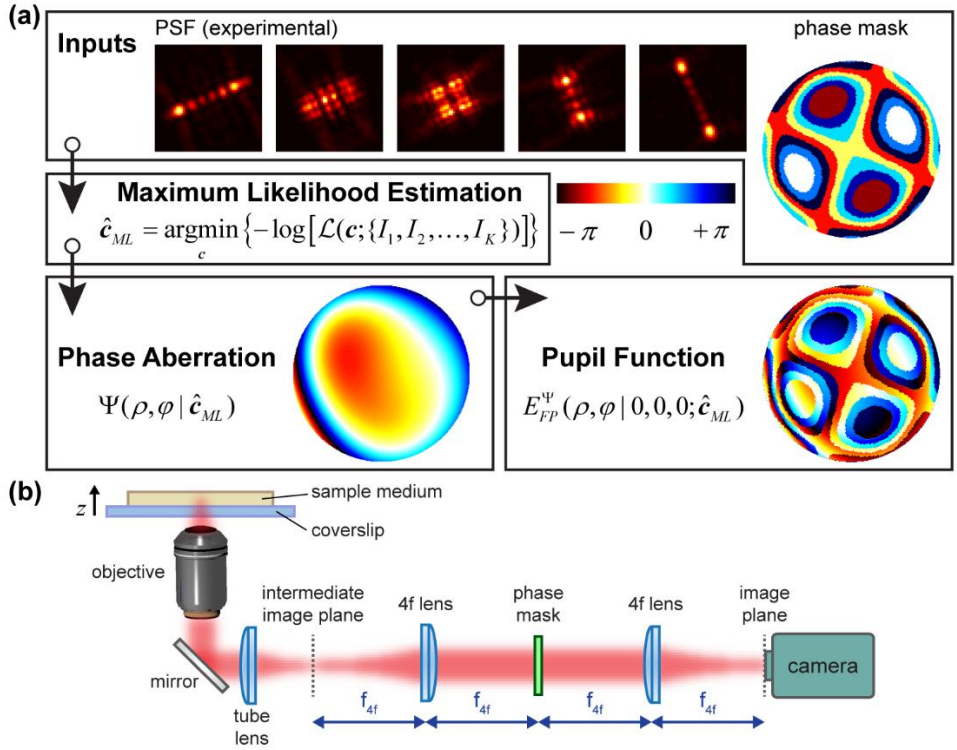

Fig. 1. (a) Schematic of the phase retrieval algorithm. A map of the phase mask pattern is used to produce the theoretical model, and a set of experimental images of the PSF is used to perform the MLE step. The estimation procedure returns a phase aberration term, which is added to the original phase mask pattern to produce the overall pupil function. In all cases, only the phase portion of the field is shown. (b) Schematic of the emission path of the microscope. Fluorescence collected from the sample by the objective is focused by a tube lens onto the intermediate image plane and relayed by a pair of 4$f$ lenses onto the final image plane, where the camera is placed. The phase mask is placed in the Fourier plane, located half way between the two 4$f$ lenses.

The Zernike polynomials form an infinite basis on the unit circle, but for computational complexity reasons we limit our parameter space to $J = 15$. Even including only 12 Zernike polynomials we find good agreement between the phase-retrieved model PSF and the experimentally observed images while minimizing the number of coefficients in order to keep the estimation problem tractable.

## 3. Phase retrieval procedure

In order to determine the coefficients of the Zernike polynomials, we use a phase retrieval method based on MLE [Fig. 1(a)]. This is an estimation technique which determines the parameters of a statistical model based on a set of experimental observations and an assumption about the noise model that underlies those observations [33]. In the context of the phase retrieval procedure considered here, the parameters to be estimated are the coefficients of the Zernike polynomials in the phase aberration term in Eq. (7), the experimental observations are images of the tetrapod PSF from a point emitter acquired at various defocus values as described in Section 5, and the noise in each pixel is assumed to be dominated by Poisson shot noise due to photon arrival statistics in a constant background.

When the data is acquired using the electron multiplication feature of an EMCCD camera, the excess noise produced by the multiplication process should also be taken into account. For modern cameras, the excess noise has the effect of reducing the quantum efficiency of the camera by a factor of two [34], which we model by dividing the observed photon counts by two prior to beginning the analysis and treating the resulting image as obeying Poisson noise statistics. This is an approximation of more detailed EMCCD noise models, which account for additional considerations such as readout noise that are more significant when modeling pixels with low photon counts [35].

### 3.1 Likelihood function

To construct the likelihood function, we consider the form of the experimental calibration scan. A set of $K$ images $\{I_1, I_2, \ldots, I_K\}$ of a single emitter is measured, each with the objective focus placed at a different position such that the displacement of the emitter from the focal plane is $z_o \in \{z_1, z_2, \ldots, z_K\}$. Each of the images consists of $S$ pixels, and we denote the measured photon count in each pixel by $\tilde{X}_{k,s}, k \in \{1, 2, \ldots, K\}, s \in \{1, 2, \ldots, S\}$.

To estimate the likelihood of the measurement $\tilde{X}_{k,s}$, we treat the photon counts in the different pixels of the image data as independent and Poisson-distributed, with a mean value given by the imaging model. We first estimate the background photon level per pixel, $b$, by averaging the photon counts per pixel in a region near the emitter. Then, total signal photons, $N$, are estimated by summing the total photons from the image at $z_o = 0$ and subtracting the total background photons, $Sb$. For a given vector of aberration coefficients, $c$, simulated images of the PSF can be produced at the positions $z_o = \{z_1, z_2, \ldots, z_K\}$, resulting in expected photon counts $X_{k,s} | c$, with $\tilde{X}_{k,s} \sim \text{Pois}(X_{k,s} | c)$ assumed based on the noise model. The likelihood function over all pixels in the data set then takes the form

$$\mathcal{L}(c; \{I_1, I_2, \ldots, I_K\}) = \prod_{k=1}^{K} \prod_{s=1}^{S} \frac{(X_{k,s} | c)^{\tilde{X}_{k,s}} \exp(-X_{k,s} | c)}{\tilde{X}_{k,s}!} \tag{9}$$

Computationally, to estimate the vector of polynomial coefficients, $\hat{c}_{ML}$, maximizing $\mathcal{L}(c; \{I_1, I_2, \ldots, I_K\})$, we minimize its negative logarithm

$$\begin{aligned}\hat{c}_{ML} &= \underset{c}{\text{argmin}} \left\{ -\log\left[\mathcal{L}(c; \{I_1, I_2, \ldots, I_K\})\right] \right\} \\ &= \underset{c}{\text{argmin}} \left\{ \sum_{k=1}^{K} \sum_{s=1}^{S} (X_{k,s} | c) - \tilde{X}_{k,s} \log(X_{k,s} | c) + \log(\tilde{X}_{k,s}!) \right\}\end{aligned} \tag{10}$$

The last term in the sum is independent of $c$, so it can be ignored in practice with no effect on the minimization process.

Once the maximum likelihood estimate of the coefficients, $\hat{c}_{ML}$, is determined, the corresponding phase aberration, $\Psi(\rho, \varphi | \hat{c}_{ML})$, and pupil function, $E_{FP}^{\Psi}(\rho, \varphi | 0, 0, 0; \hat{c}_{ML})$, can be computed. From the pupil function, an image for any set of parameters $\theta$ can be determined,

so the phase-retrieved PSF model is complete. The algorithm is summarized graphically in Fig. 1(a).

### 3.2 Additional estimation parameters

Although the term $\Psi(\rho,\varphi|\boldsymbol{c})$ can account for arbitrarily large and complicated wavefront aberrations with large enough $J$, it is a straightforward matter to include other parameters in the estimation procedure to improve the resulting PSF model.

In addition to the physically-motivated improvements to the theoretical imaging model described in Section 2.2, we include additional estimation parameters, $\Delta\xi$ and $\Delta\eta$, corresponding to lateral misalignment of the phase mask. These are incorporated into the imaging model as lateral displacements of the tetrapod pattern by performing the transformation

$$\mathfrak{M}(\xi',\eta') = \mathfrak{M}(\xi+\Delta\xi, \eta+\Delta\eta) \qquad (11)$$

and using a new vector, $\boldsymbol{d} = [c_4, c_5, ..., c_J, \Delta\xi, \Delta\eta]$, in place of $\boldsymbol{c}$ in Eq. (10) to determine the maximum likelihood estimate, $\hat{\boldsymbol{d}}_{ML}$. Although this involves the estimation of two additional parameters, it is a useful feature to account for the finite precision of the phase mask alignment procedure or the slight lateral offset of the phase mask that can occur over time. In our system, typical lateral phase mask misalignments in the Fourier plane were estimated to be about 10-30 µm, which corresponds to about 1% of the mask diameter. This slight lateral shift is a minor aberration compared with the contribution from $\Psi(\rho,\varphi|\boldsymbol{c})$ but contributes a different type of phase function than the smooth Zernike polynomial-based term due to the discontinuities in phase present in the fabricated mask.

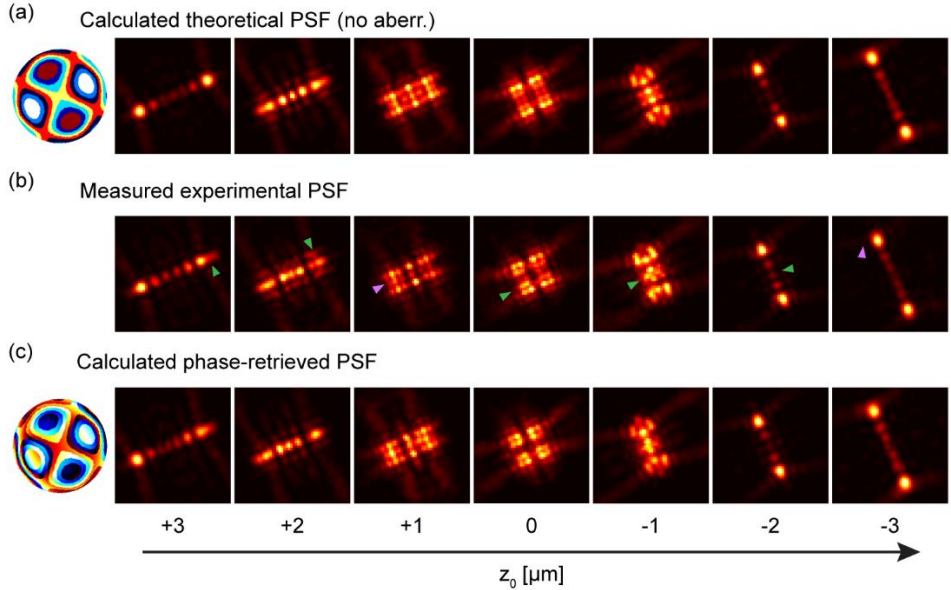

Fig. 2. (a) Left: phase-only portion of the theoretical pupil function with the tetrapod phase mask. Right: tetrapod PSFs calculated using the phase pattern on the left. (b) Experimentally-acquired images of the tetrapod PSF. Green (purple) arrows indicate features reproduced better in the phase-retrieved (calculated theoretical) PSF model, but a more quantitative comparison is shown in Fig. 3. (c) Left: phase-only portion of the phase-retrieved pupil function. Right: tetrapod PSFs calculated using the phase pattern on the left. Side length is 6 µm in all PSF images.

Additional estimation parameters motivated by other physical features of the microscope setup can be added, such as a rotation angle, axial misalignment, or tilt of the phase mask away from the optical axis, or an effective numerical aperture or magnification of the system deviating from the design value. Notably, while these parameters can be added to the imaging model in a similar fashion to the lateral misalignment of the mask, additional parameters can also complicate the optimization process, so it is important to judiciously parameterize the imaging model. Here, consideration of additional parameters was limited to lateral shifts of the phase mask.

## 4. Results and discussion

### 4.1 Phase retrieval results

To test the performance of the phase retrieval algorithm, a set of images of a fluorescent bead was acquired as described in Section 5. The theoretical pupil function for the system was calculated from the digital model of our fabricated tetrapod phase mask, and the phase-retrieved pupil function was determined from the experimental data using the algorithm described in Section 3. Sample images of the theoretical, experimental, and phase-retrieved PSFs are shown in Fig. 2. Qualitatively, all three PSFs look similar since the total phase aberration in our microscope is small, especially in comparison to the contribution of the tetrapod phase pattern, $\mathfrak{M}(\rho,\varphi)$. However, some minor differences between the three sets of images are observed.

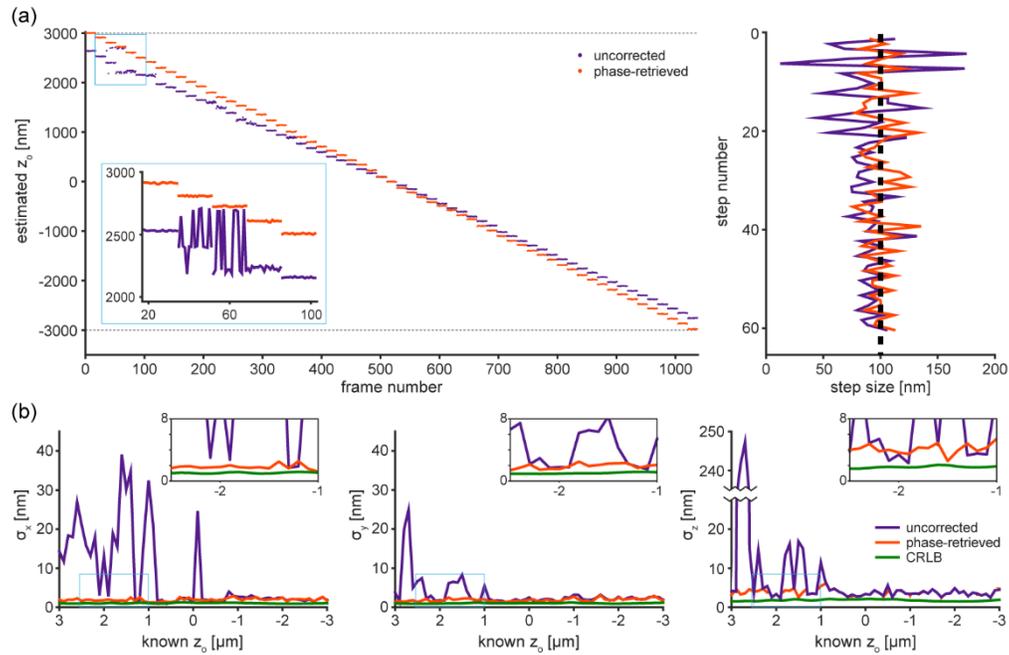

Fig. 3. Results of axial scan of a fluorescent bead. (a) Left: estimated axial position at each frame in the scan. Each fit is shown as a dot, and dashed lines are added at ±3000 nm to guide the eyes. Inset shows a close-up of five steps, with localizations in each axial step connected by lines to illustrate fluctuations. Estimates of $z_o$ are arbitrarily defined to be 0 in the middle of both scans. Right: step size plotted against axial step number. The true value is indicated with a dashed line. (b) Localization precision in x (left), y (middle), and z (right) as a function of known z position for the uncorrected (theoretical) and phase-retrieved PSF. Insets show magnified views of the curves over part of the axial range, indicated by blue boxes in the main plots. Localization precision is determined from the full set of localizations at each axial position. CRLB is calculated for the phase-retrieved PSF with 95000 signal photons and 3.7 background photons per pixel, as measured for the bead.

In order to quantitatively compare the performance of the uncorrected theoretical and phase-retrieved PSF models, we performed scans of fluorescent beads using a step size of 100 nm and localized the resulting PSFs using maximum likelihood estimation of the PSF parameters $\theta$ as described in Section 5. An example of the results, shown in Fig. 3, demonstrates the necessity of the phase retrieval procedure. When the theoretical PSF model is used to localize the bead, mismatch between it and the experimental PSF leads to mislocalizations [Fig. 3(a)], degeneracies in the likelihood function between multiple axial positions [Fig. 3(a), inset], and poor localization precision in some regions of the scan [Fig. 3(b)]. In addition, poor precision and accuracy in the axial step size is observed throughout the scan. The poor performance for

$z_o$<0 may be due to the design of the objective itself, which was not intended to be focused within the coverslip.

By contrast, the phase-retrieved model demonstrates significant improvement over the theoretical model, offering high precision and accuracy over the entire axial range. Calculation of the CRLB (see Section 5.5) reveals that the phase-retrieved model is close to the theoretical limit of localization precision in all three dimensions. As shown in Fig. 3(b), the mean CRLB is ~1 nm in the lateral dimensions and ~2 nm in the axial dimension, while the mean localization precision achieved by the phase-retrieved model is ~2 nm in the lateral dimensions and ~4 nm in the axial dimension.

The slight discrepancies between the phase-retrieved and CRLB can be attributed to outstanding mismatch between the model and experiment. First, the CRLB calculation assumes a Poisson noise model within each pixel, which only approximates the true behavior of the EMCCD due to the possible presence of additional noise sources [35]. Additionally, the assumptions in the CRLB calculation (see Section 5.5) of a constant background level in each pixel and a constant total signal in each frame are violated to some extent in any real measurement. Finally, another contribution to discrepancies between the expected and calculated localization precision comes from differences between the PSF shapes that remain despite the addition of the phase aberration term to the model. These may be due to using only a small set of Zernike polynomials, treating the emitter as monochromatic, or approximating the vectorial nature of light by a scalar field.

The precision which can be achieved when localizing any emitter depends significantly on signal and background photon levels. The demonstration here was performed in the high-photon limit so as to minimize the contributions of photon shot noise and emphasize outstanding mismatch between the model and experimental PSFs. Naturally, one can expect a lower signal-to-noise ratio in experiments with dimmer emitters such as single molecules. As the signal-to-noise ratio is decreased, its contribution to degraded localization precision will dominate the contribution from model mismatch shown here.

*4.2 Reproducibility*

In addition to enhancing the performance of the localization procedure and allowing for the full 6 µm axial range of the tetrapod PSF to be utilized, the phase retrieval algorithm outputs an estimate of the phase aberration present in our imaging system. This estimate serves as a qualitative indicator of the magnitude and shape of the phase aberration, as well as a quantitative measure of its functional form which can be used to inform corrective procedures such as the introduction of adaptive optics into the imaging system.

In order to characterize the precision with which we are able to estimate the coefficients of the Zernike polynomials present in our phase aberration, we performed phase retrieval using scans of 35 different fluorescent beads as described in Section 3. The emission from each bead was spectrally separated into two color channels (termed "green" and "red") using a dichroic mirror, resulting in two scans at slightly different wavelengths, using two different dielectric tetrapod phase masks. The use of two spectral channels is purely for illustration that the entire procedure can easily handle such a situation. From each scan, a set of 12 coefficients was obtained, each corresponding to the estimated contribution of a different Zernike polynomial to the total phase aberration. The distribution of phase aberrations in each spectral channel and for each Zernike polynomial is shown in Fig. 4.

The dominant phase aberration observed in both channels was the defocus mode (Noll index 4). This is because, in experiments, the tetrapod PSF is manually identified as "in focus" ( $z_o = 0$ ) when the observed image consists approximately of four squares, as shown in the middle panel of Fig. 2(b). However, the theoretical models based on the fabricated tetrapod phase masks do not produce such an image for $z_o = 0$, but rather for an axial offset of a few hundred nanometers. As a result, the phase retrieval algorithm must compensate for this offset by adding a contribution from the available radially-symmetric Zernike polynomials: the defocus mode and, to a lesser extent, the spherical aberration mode (Noll index 11). Moreover, the large standard deviation in the estimated coefficient of the defocus mode can be attributed to slight axial offsets between the PSFs treated as "in focus" during the different scans. The measured standard deviation of the defocus mode was ~0.3 radians, which suggests a fluctuation between scans of approximately 40 nm in the location along the optical axis manually selected as $z_o = 0$. Importantly, the remaining modes were identified with generally high precision, and the contribution of the aberrations was seen to decay substantially with increasing Noll index, which justifies the truncation of the phase aberration term in Eq. (7) at $J = 15$. On the other hand, expanding the basis to $J > 15$ could be done if needed, at the cost of increased computational complexity in the optimization.

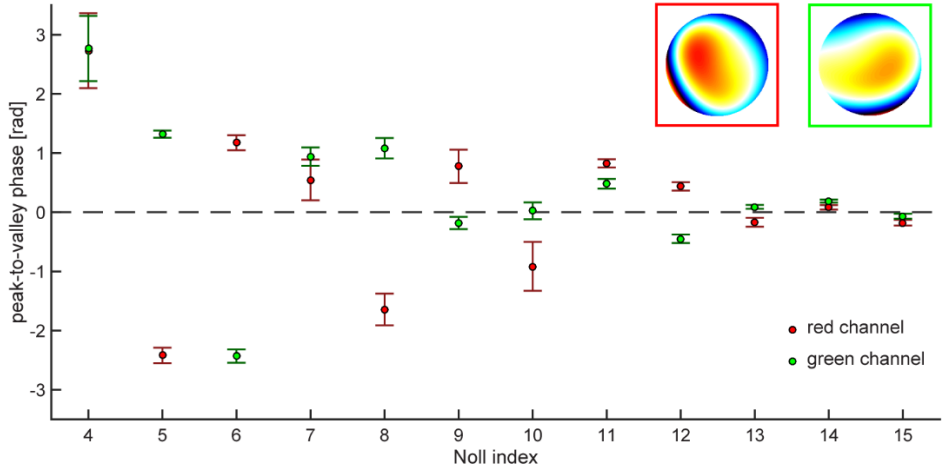

Fig. 4. Reproducibility from 35 scans (per channel). Circles show mean Zernike polynomial peak-to-valley phase difference; error bars show standard deviation. Negative peak-to-valley values correspond to negative polynomial coefficients. Mean signal photons across all scans were 54000 (green) and 46000 (red). Insets show phase aberrations calculated from mean value of each aberration.

Finally, we note that the shape of the total aberration in the two channels (Fig. 4, insets) looks similar up to a reflection about the line $\xi = \eta$. This reflection is present in the two-color microscope setup [11], suggesting that the dominant aberrations in the imaging system are due to optics placed before the splitting of the spectral channels by the dichroic mirror. The dichroic mirror splits the optical path into two (one path for each spectral channel), so that, whereas optics placed upstream of the dichroic mirror contribute to aberrations in both spectral channels, optics downstream of the dichroic mirror (namely, the phase mask, second 4*f* lens, and knife edge mirror as shown in [11]) contribute only to aberrations in the spectral channel in which they are placed. As a result, the slight differences between the aberrations observed in the two channels, after consideration of the reflection and the difference in wavelength, can be attributed to the downstream optics.

## 5. Experimental methods

### 5.1 Sample preparation and optical setup

Fluorescent beads (Molecular Probes, F8800, orange fluorescent, 540/560) with a diameter of 100 nm were diluted to a final concentration of ~6 pM in 1% (w/w) polyvinyl alcohol in water. The solution was spin-coated onto a plasma-etched glass coverslip, which was then mounted on an inverted Nikon TE300 microscope altered to provide a two-color engineered PSF as described in [36]. A small chamber (Grace Bio-Labs, SecureSeal Hybridization Chamber) was added on top of the coverslip and filled with immersion oil ($n = 1.518$, Zeiss, Immersol 518F) to provide index-matching. The sample was pumped by a 561 nm laser (Coherent Sapphire) in the epi-illumination configuration with an intensity of ~0.1 kW/cm$^2$. Fluorescence was collected by an oil-immersion objective (Olympus, PLAPON60XOSC2 60X/1.4) and passed through a dichroic filter (Chroma, ZT405/488/561rdc), a 561 nm notch filter (Semrock, NF03-561E), and a 523/610 dual bandpass filter (Semrock, FF01-523/610). The internal tube lens of the microscope was removed and replaced with an external doublet lens (*f* = 400 mm), which focused the fluorescence onto the intermediate image plane of the system. The image was then split into two spectral channels at 560 nm by a dichroic beamsplitter (Semrock, FF560-FDi01), and each channel was relayed onto different regions of an EMCCD (electron-multiplying charge-coupled device) camera (Andor iXon+) by a pair of *4f* lenses (*f* = 120 mm). Custom fabricated 6 µm tetrapod quartz phase masks with design wavelengths of 660 nm (red) and 550 nm (green) were placed in the Fourier plane of each spectral channel to create the tetrapod PSFs. An additional 605 nm long pass filter (Chroma, HQ605LP) was placed in the red channel.

## *5.2 Scanning procedure*

The sample was imaged continuously at a frame rate of 20 Hz while the objective was scanned along the optical (*z*) axis from +3.5 µm to -3.5 µm using a piezoelectric objective scanner (Mad City Labs, C-Focus). A step size of 250 nm was used for phase retrieval data and a step size of 100 nm was used for localization studies. In each case, 20 frames were acquired at each step and an electron multiplication gain of 100 was used.

## *5.3 Localization procedure*

Localization of tetrapod PSFs was performed using a maximum-likelihood estimator with a Poisson shot noise model, analogous to the one described in Section 3.1, but parameterized by the five terms uniquely determining a PSF: 3D emitter position $(\hat{x}, \hat{y}, \hat{z})$, total signal photons $\hat{N}$, and background photons per pixel $\hat{b}$.

Theoretical model PSFs were generated by setting the wavefront aberration term in Eq. (8) to zero in the imaging model, while phase-retrieved model PSFs used the aberration term as determined by the phase retrieval algorithm. In each case, the models were generated at 250 nm intervals throughout the 6 µm range of the tetrapod PSF and a slight Gaussian blur was applied to each model image to account for the finite size of the fluorescent bead. Once the appropriate coarse $z$ position within this library was identified, a cubic 3D interpolation of the model PSF between these intervals was used to rapidly search the $z$ coordinate for the maximum likelihood estimate, $\hat{z}$.

## *5.4 Calibration and correction of localization biases*

After a maximum likelihood estimate is obtained, an additional step is taken to correct for localization biases in each spatial dimension. These biases are defocus-dependent offsets in position estimates which are caused by certain low-symmetry aberrations and discrepancies between the data and the model PSFs [37]. Similar effects occur for other 3D PSFs, such as for the DH-PSF [38]. Although the phase retrieval procedure greatly reduces the apparent biases relative to those observed when using the theoretical PSF model, we perform an additional calibration procedure after the phase retrieval algorithm has been completed and $\hat{c}_{ML}$ has been determined, the purpose of which is to compensate for residual biases that exist due to outstanding mismatch between the model and experimental PSFs.

We begin with the assumption that, during the scan used for the phase retrieval procedure, the lateral position of the bead is $(x_o, y_o) = (0, 0)$ and the axial position takes on a known value $z_o \in \{z_1, z_2, \ldots, z_K\}$. From MLE fits of the images from the scan as described above, we obtain a position estimate $(\hat{x}_i, \hat{y}_i, \hat{z}_i)$ for each $i \in \{1, 2, \ldots, K\}$. Using a polynomial fit, in each spatial dimension a bias function is acquired which describes the offset of the MLE fit from the known value as a function of the known axial position. By inverting this function for the axial coordinate, we calculate the true axial position, $z_o$, from the estimate, $\hat{z}_i$. Then, the lateral bias terms are determined via the other polynomial fits using the calculated $z_o$.

Once the polynomial fits are determined based on the calibration scan corresponding to a given pupil function, they are used to correct all MLE fits performed with the corresponding PSF model. Since the polynomials are generally slowly-varying in axial position, they have minimal effects on localization precision while enforcing accuracy under the assumptions made about the bead positions during the calibration scan.

*5.5 Calculation of Cramér-Rao lower bounds*

In order to benchmark the localization precision, the best-case precision was determined via calculation of the Cramér-Rao lower bound (CRLB) in each axis (x, y, and z) as a function of axial emitter position. The CRLB bounds the variance of any unbiased estimator from below to the inverse of the Fisher information, given a PSF and a noise model.

Here, the Fisher information is described by a 5x5 matrix $\mathbb{I}(\boldsymbol{\theta})$ which describes the sensitivity of the measured PSF to the five estimation parameters in the vector $\boldsymbol{\theta}$. The elements of the matrix are defined by

$$\mathbb{I}_{ij}(\boldsymbol{\theta}) = -\mathbb{E}\left[\frac{\partial^2 \ln f(\xi;\boldsymbol{\theta})}{\partial \theta_i \partial \theta_j}\right] \quad (12)$$

where $f(\xi;\boldsymbol{\theta})$ is the probability of measuring signal $\xi$ given the underlying parameters $\boldsymbol{\theta}$, $\theta_i$ and $\theta_j$ are parameters in $\boldsymbol{\theta}$, and $\mathbb{E}[\cdot]$ is the expected value operation. In the case presented here, the signal $\xi$ is the measured PSF which is assumed to be subject to pixelation and Poisson noise statistics. An additive Poisson-distributed background with mean value $b$ in each pixel is also added. Under these assumptions, the Fisher information matrix elements are given by [23]

$$\mathbb{I}_{ij}(\boldsymbol{\theta}) = \sum_{s=1}^{S} \frac{1}{\mu(s|\boldsymbol{\theta})+b}\left[\frac{\partial \mu(s|\boldsymbol{\theta})}{\partial \theta_i}\right]\left[\frac{\partial \mu(s|\boldsymbol{\theta})}{\partial \theta_j}\right] \quad (13)$$

where $\mu(s|\boldsymbol{\theta})$ is the expected value of the model PSF in pixel $s$ and the sum is over all $S$ pixels in the image. The CRLB vector is given by the diagonal of the inverse of the Fisher information matrix, such that the precision with which each parameter in $\boldsymbol{\theta}$ can be estimated is given by

$$\sigma_i^2 \geq \left\{\left[\mathbb{I}(\boldsymbol{\theta})\right]^{-1}\right\}_{ii} \equiv CRLB_i \quad (14)$$

Fig. 3(b) shows the localization precision $\sigma_i = \sqrt{CRLB_i}$ calculated in each spatial dimension ($i \in \{x, y, z\}$) over the 6 µm axial range of the tetrapod PSF using the phase-retrieved model.

*5.6 Phase retrieval code*

The phase retrieval and PSF localization procedures discussed in this paper were performed using custom MATLAB code. A publicly-available version of the phase retrieval code is available in Code File 1 (Ref. [39]).

## 6. Conclusion

A simple, global phase retrieval method has been presented to estimate the optical aberrations in a microscope and incorporate them into the imaging model in order to perform single-emitter localization for complex PSF designs. The method only requires a set of calibration images of an isolated emitter and a model of the phase mask placed in the Fourier plane.

We have shown that this method can be used to perform single-emitter localizations with precision that approaches the CRLB in all three dimensions over the entire 6 µm axial range of the tetrapod PSF. We have also shown the utility of the phase retrieval procedure in determining the unique aberrations for two spectral channels separately.

A key advantage of this method is that the phase retrieval is carried out in the presence of the phase mask, so that aberrations due to misalignment or inaccuracies in the fabrication of the optic can be incorporated into the imaging model either explicitly as optimization parameters (as in the case of the lateral misalignment terms described in Section 3.2) or implicitly as a contribution to the sum of Zernike polynomials in Eq. (7). Alternatively, the aberrations estimated by this procedure can be provided to an adaptive optical element such as a deformable mirror in order to improve the image quality experimentally.

This phase retrieval algorithm can also easily be generalized to include different optimization terms that may be relevant to estimating the pupil function of the microscope. For instance, although we have demonstrated this procedure in the context of the 6 µm tetrapod PSF, the framework is general and can readily be adapted to the phase retrieval of pupil functions in microscopes with other phase-engineered PSFs, requiring only that an appropriate model of the phase mask be selected.

Finally, although the theoretical pupil function presented here was derived for the case of scalar diffraction and an index-matched sample, in principle it can be extended to a vectorial diffraction model or mismatched refractive indices by the inclusion of additional terms in the pupil function.

## Funding

National Institute of General Medical Sciences (NIGMS) (R35GM118067); National Institute of Biomedical Imaging and Bioengineering (U01EB021237).

## Acknowledgments

We thank Maurice Lee and Anna-Karin Gustavsson for valuable discussions. We are also very grateful to Maurice Lee for providing the fabricated tetrapod phase masks. Y.S. is a CAC Fellow.